\documentclass[twocolumn,twoside,slac_two]{revtex4}
\usepackage{graphicx}
\usepackage{fancyhdr}
\begin{document}

\title{OVRO 40m blazar monitoring program: Understanding the relationship between 15 GHz radio variability properties and gamma-ray activity in blazars}

%

\author{W. Max-Moerbeck, J.L. Richards, V. Pavlidou, T. Hovatta, O.G. King, T.J. Pearson, A.C.S. Readhead, R. Reeves, M.C. Shepherd}
\affiliation{California Institute of Technology, Pasadena, CA, USA}
%

\begin{abstract}
A large sample of known and likely gamma-ray blazars has been monitored twice per week since late 2007 at 15 GHz with the Owens Valley Radio Observatory (OVRO) 40-meter Telescope. The sample contains about 1700 sources, including the initial sample of 1158 sources above declination $-20^{\circ}$ from the Candidate Gamma-Ray Blazar Survey (CGRaBS) plus all the blazars associated with \emph{Fermi}-LAT detections as released in the \emph{Fermi} AGN catalogs. Using statistical likelihood analyses, we compare the variability amplitude for various sub-populations within our sample. These include comparisons of gamma-ray-loud versus quiet objects, BL Lac objects versus flat-spectrum radio quasars, and a study of the variability amplitude trend with redshift. To learn about the location of the gamma-ray emission region we study the significance of peaks in the radio/gamma-ray cross-correlation using Monte Carlo simulations. First results for 52 sources with data from both the high-confidence \emph{Fermi} Large Area Telescope Bright AGN Sample and the first 2 years of our monitoring program are presented.  We find that assuming a power spectral density with power law slope of -2 at 15 GHz and -1.5 at gamma-ray energies, 7 of our objects show cross-correlations at the $3\sigma$ level. ÊWe are now studying the physical significance of these correlations by further exploring the range of power law slopes that are consistent with the data. An extension of this to a larger sample and longer light curves is underway and preliminary results are presented. We also describe KuPol, the new digital Ku-band receiver being constructed for the 40-meter telescope. This new receiver will provide total intensity and linear polarization measurements over the 12-18 GHz band, with 16 MHz spectral resolution. The polarization data will provide important clues about the magnetic field configuration in the radio emission region.
\bigskip
\bigskip
\bigskip
\bigskip
\end{abstract}

\maketitle

\thispagestyle{fancy}

\section{INTRODUCTION}
A long standing question in blazar astrophysics is the location of the gamma-ray emission site. Two alternatives are that the high-energy emission is produced close to the base of the jet \citep[][]{blandford_and_levinson_1995} or further out in the jet from the same shocked regions that are responsible for the radio emission seen in VLBI observations \citep[][]{jorstad_2001}. VLBI observations can resolve the jets of blazars on scales that show moving emission features. The same is not possible at higher energies so another approach has to be taken in order to locate the site of the high-energy emission. Even though \emph{Fermi} provides only limited angular resolution, it provides continuous  monitoring of the whole sky, a feature that can be exploited to help locate the site of the gamma-ray emission through the study of correlated variability. Seeing this opportunity we started a radio monitoring program of candidate gamma-ray blazars in mid 2007. The candidate gamma-ray blazars were selected to be similar to EGRET-detected ones \citep[][]{healey_2008}. New \emph{Fermi}-detected sources based on first and second year active galactic nuclei \emph{Fermi} catalogs \citep[][]{abdo_1lac_2010, ackermann_2lac_2011} have been added to the monitoring, and the current sample contains more than 1700 sources.

A number of complementary multi-wavelength monitoring programs are underway, monitoring different samples of sources at different wavebands and cadences. Our strategy is to monitor the largest sample of radio sources. These are monitored continuously, independent of the activity at other wavelengths or gamma-ray detection, thus allowing for an unbiased view of their activity, and for a characterization of gamma-ray detected versus non- detected sources. Another important element is the use of robust statistical tools that allow for a quantitive assessment of the relationship between different wavelengths. In order to enhance our monitoring efforts at radio wavelengths we are also building a new receiver for the Owens Valley Radio Observatory (OVRO) 40 meter telescope that will enable polarization monitoring and double the bandwidth.

Here we present a brief overview of our radio program, an examination of the variations of radio variability properties over different source populations, and a study of correlated radio and gamma-ray variability using publicly available \emph{Fermi}-LAT light curves for the brightest gamma-ray sources detected in the first three months of \emph{Fermi} operations. Preliminary results on the variation of the radio power spectral density are also included. A brief description of the capabilities of a new receiver under construction is also presented.

\section{OVRO 40 METER TELESCOPE BLAZAR MONITORING PROGRAM}

Sources have been monitored using the Owens Valley Radio Observatory 40 meter Telescope since mid 2007. The sample contains all the sources included in  the ``Candidate Gamma-Ray Blazar Sample" (CGRaBS) \citep[][]{healey_2008} north of declination $-20$ degrees plus all the sources included in the first catalog of \emph{Fermi} detected AGN (1LAC) \citep[][]{abdo_1lac_2010} and more recently the second catalog of \emph{Fermi} detected AGN (2LAC) \citep[][]{ackermann_2lac_2011}.

We monitor their flux density in a 3 GHz band centered at 15 GHz and obtain two measurements per week for each source with a typical flux error of 4 mJy and $\sim$3\% uncertainty from pointing errors and other systematic effects. Relative calibration is obtained by measuring a noise diode signal more often than hourly and at similar elevation. The overall flux density scale is obtained from 3C 286 which is observed daily along with other calibrators that are used to check the stability of the receiver. A detailed discussion of the observing strategy and calibration procedure can be found in \citet[][]{richards_2011}.
The results of the radio variability properties are obtained using 2 years of radio monitoring from January 1, 2008 to December 31, 2009. The gamma-ray data are the light curves published in \citet[][]{abdo_variability_2010} which correspond to the sources detected in the first three months of \emph{Fermi} operations with a time duration of 11 months using weekly time binning.

\section{RADIO VARIABILITY OF BLAZARS}

\subsection{First 2 Years of Observations}
Among the defining characteristics of blazars is their variability from radio to gamma-rays. A complete study of their properties using large samples can reveal the features that make some radio blazars gamma-ray emitters. Even for the best cases in our sample the light curves have a time sampling which is uneven and often has gaps. All these are challenges that variability studies must overcome by the use of robust statistical techniques. In order to improve the robustness of the variability characterization and to obtain reliable estimates of the measurement errors in the estimated variability, we have introduced a new variability statistic which we call the intrinsic modulation index. The intrinsic modulation index is defined as $\bar{m} = \sigma_{0}/S_{0}$, where $\sigma_{0}$ is the intrinsic standard deviation of the distribution of source flux densities in time and $S_{0}$ the intrinsic source mean flux density. The best estimate of the intrinsic modulation index and error are obtained by the use of maximum likelihood analysis that assumes that the intrinsic flux values are from a Gaussian distribution and that the measurement process adds noise which is  also from a Gaussian distribution. The details of the method and the results are given in \citet[][]{richards_2011}, only a brief summary of the most important conclusions is given below.

By studying the distribution of the intrinsic modulation index we find that sources that have been detected by \emph{Fermi} have on average larger intrinsic modulation index than the non-detected sources. A similar result is found when dividing the sources between BL Lacs and FSRQs. In this case BL Lac sources are found to have larger intrinsic modulation indices than the FSRQs. Finally, a trend with redshift in which the average intrinsic modulation index decreases with increasing redshift within the FSRQ sample is found. While it would be exciting were this indicative of a cosmological evolution, a number of effects can cause such an effect (e.g., shorter rest-frame observation intervals at higher redshift) or dilute (e.g., lower rest-frame frequency at higher redshift) such a trend, so no firm conclusion can be drawn from this observation

\subsection{Updated Results with 3.5 Years of Data}
The findings described in the previous section have been revisited using an extended, 3.5 years data set. In this case we find that gamma-ray detected sources are still significantly more variable than non-detected ones. The case for the BL Lacs versus FSRQs gives different results when different parent populations are considered. For the CGRaBS sample, the BL Lacs are still found to be more radio variable than the FSRQs with $>\,3.5\sigma$ significance. For the First LAT AGN Catalog (1LAC) sample, however, FSRQs are found more radio variable than BL Lacs, albeit with less than $2\sigma$ significance. The redshift trend is still there but at reduced  significance, suggesting that this is not a real effect. The details of these results will be presented in a paper in preparation [Richards et al, in preparation].

\section{CORRELATED RADIO AND GAMMA-RAY VARIABILITY}

The results described above refer to properties that do not use the time information available in the light curves. The next step in our investigations is to include the time information which can provide constraints on the location of the gamma-ray emission site. The existence of time lags between the radio and gamma-ray emission can be investigated by studying the cross-correlation of the blazar light curves at these two different wavelengths. Even though \emph{Fermi} and our radio monitoring program are observing the sources on a regular basis we do not obtain completely uniform time coverage of the light curves. Many effects contribute to make the time sampling uneven, for example radio observations cannot be made during poor weather, while limited gamma-ray sensitivity results in non-detections during periods of low emission. Methods that can use these particular data sets and provide quantitative information on the statistical significance of the cross-correlations are required but have only been applied in a few cases.
A description of some of the techniques we have been using and the results we find are given below.

\subsection{Estimating the Significance of the Cross-correlations}

\begin{figure}[!h]
\centering
\includegraphics[width=8.5cm]{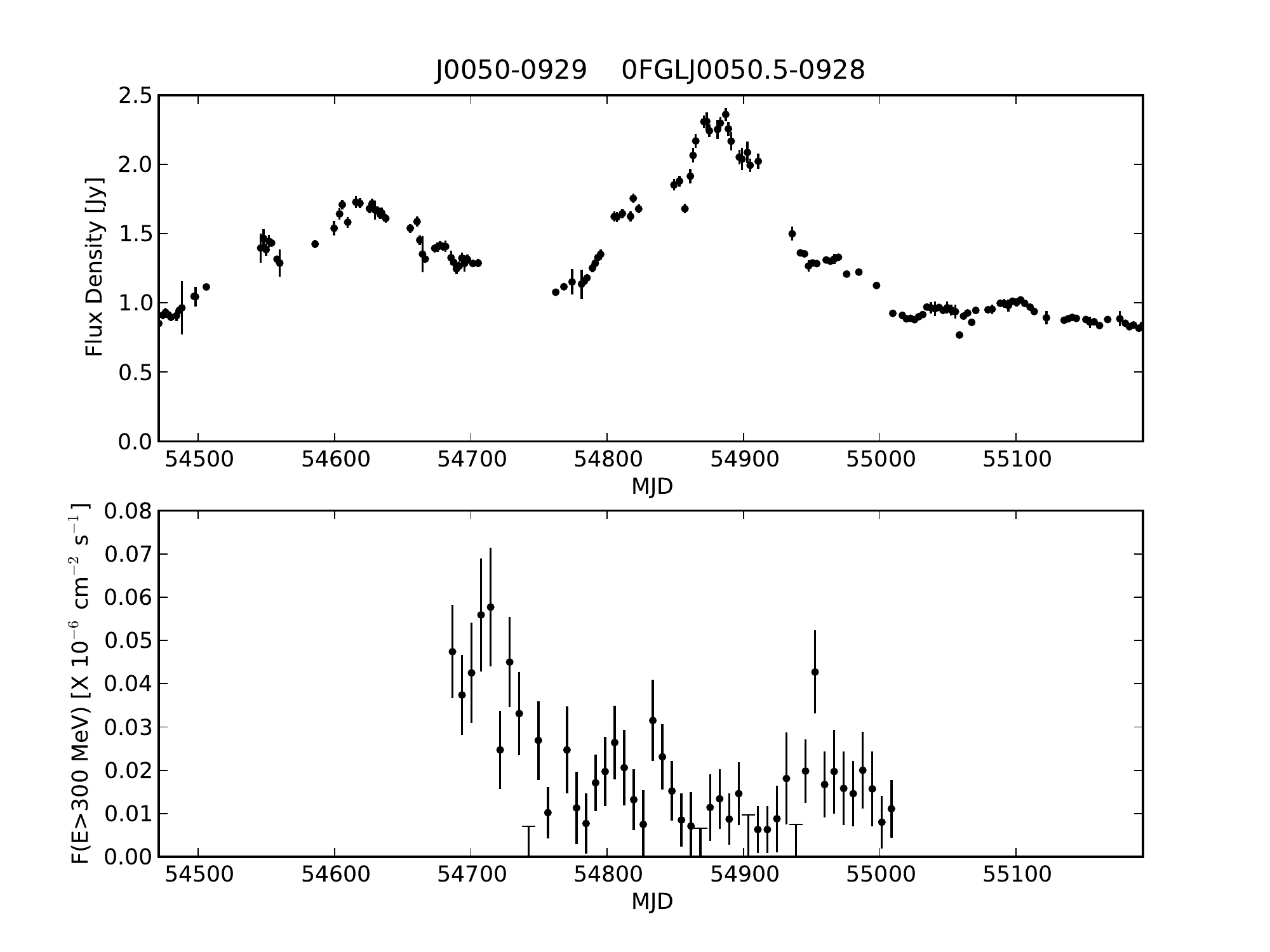}
\includegraphics[width=8.5cm]{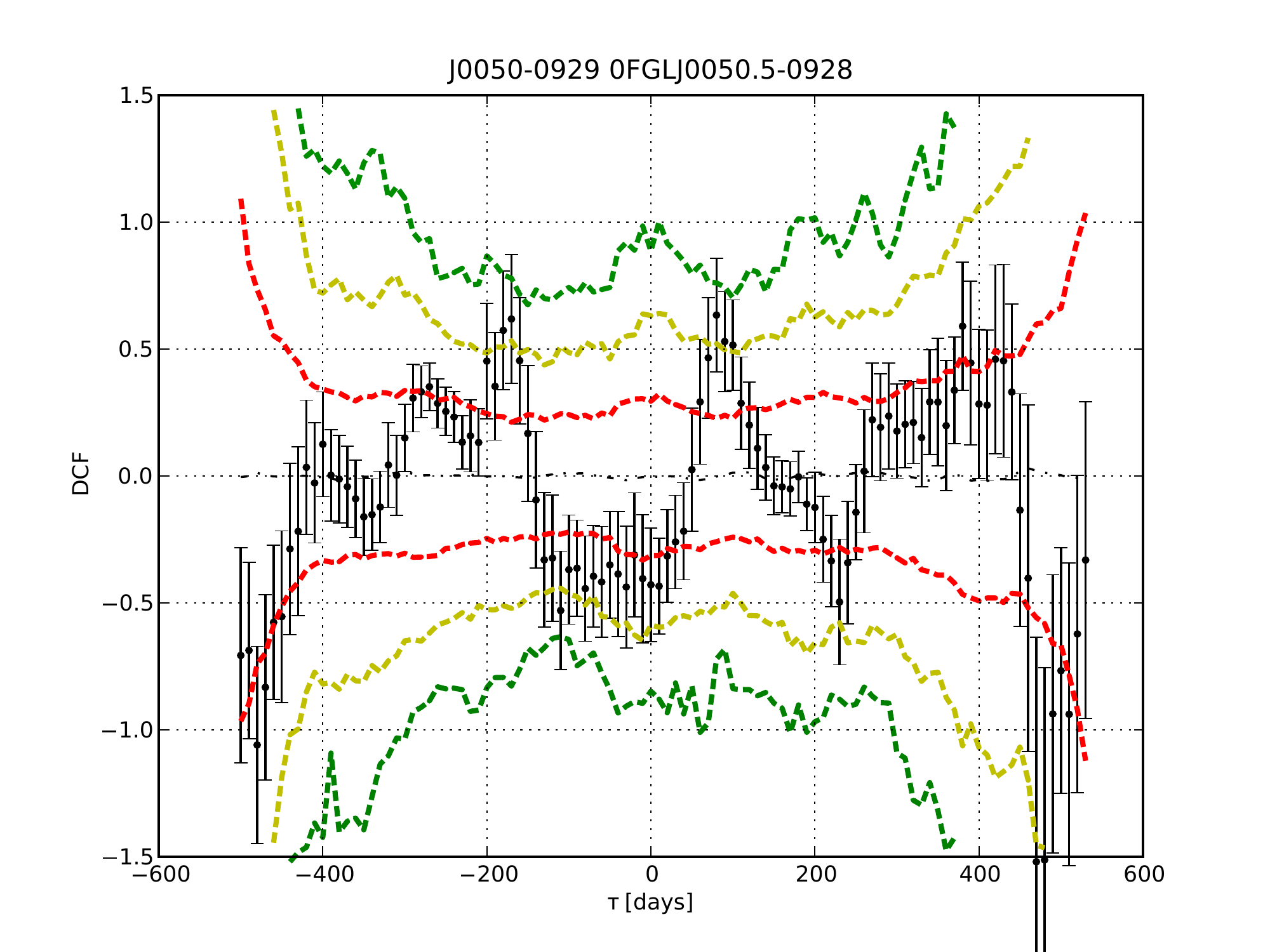}
\caption{Example of the significance of the cross-correlation using $\beta_{radio} = 2.0$ and $\beta_{\gamma} = 1.5$ for the simulated light curves. The black dots represent the DCF for the data, while the color contours the distribution of random cross-correlations obtained by the Monte Carlo simulation (red: $1\sigma$, yellow: $2\sigma$ and green: $3\sigma$). In this particular case, none of the cross-correlation peaks is above the $3\sigma$ level required for a significant detection of cross-correlation.} \label{xcorrsigexample}
\end{figure}

Methods for cross-correlating unevenly sampled time series have been developed and widely used by the reverberation mapping community \citep[e.g.][]{peterson_1993}.  Because we have a similar problem, we have adopted the \citet[][]{edelson_1988} prescription that provides an estimate of the cross-correlation between two light curves, called the discrete cross-correlation function (DCF). This method does not provide an estimate of the significance of the cross-correlations, and for that purpose we have used Monte Carlo simulations that assume the power spectral densities of the light curves can be approximated as simple power laws (PSD $\propto 1/f^{\beta}$).  Related methods have been proposed and used in the past \citep[e.g.][]{edelson_1995, uttley_2003, arevalo_2008, chatterjee_2008}. The basic idea of the method is to simulate a large number of independent light curve pairs that replicate the sampling and measurement error distribution of the original data. Using the simulations, distributions of the random discrete cross-correlation function for each time lag are obtained. These distributions provide an estimate of the probability of obtaining a high value of the correlation by a random association of two unrelated flares. A detailed description of the method will be provided in Max-Moerbeck et al, in preparation.
The main assumption of the method is that a simple power law power spectral density can be used to describe the variability properties of blazar light curves at radio and gamma-rays. This seems to be the case at radio frequencies \citep[][]{chatterjee_2008, hufnagel_1992} and gamma-rays \citep[][]{abdo_variability_2010}. Both papers have used limited number of sources and show that this is indeed a plausible model. As a trial of the method, we have estimated the significance of the radio and gamma-ray cross-correlations using values for the PSD power law slopes of -2 and -1.5 for the radio and gamma-ray variability respectively, which are consistent with the literature cited above. An example is shown in Fig. \ref{xcorrsigexample}.

Application to the brightest \emph{Fermi} sources detected in the first 3 months \citep[][]{abdo_variability_2010} with data in our monitoring program show that 7 out of 52 sources have cross-correlations at the $3\sigma$ level. It remains to be seen how physically significant these correlations are once we have explored the effects of varying the power law slopes. The time lags we found cover a broad range of values and no clear trend is seen in this sample. A complete description of the Monte Carlo simulations along with this first results will be presented in a paper in preparation [Max-Moerbeck et al. in preparation].

\subsection{Power Spectral Density for the Radio Light Curves}

\begin{figure}[!h]
\centering
\includegraphics[width=8cm]{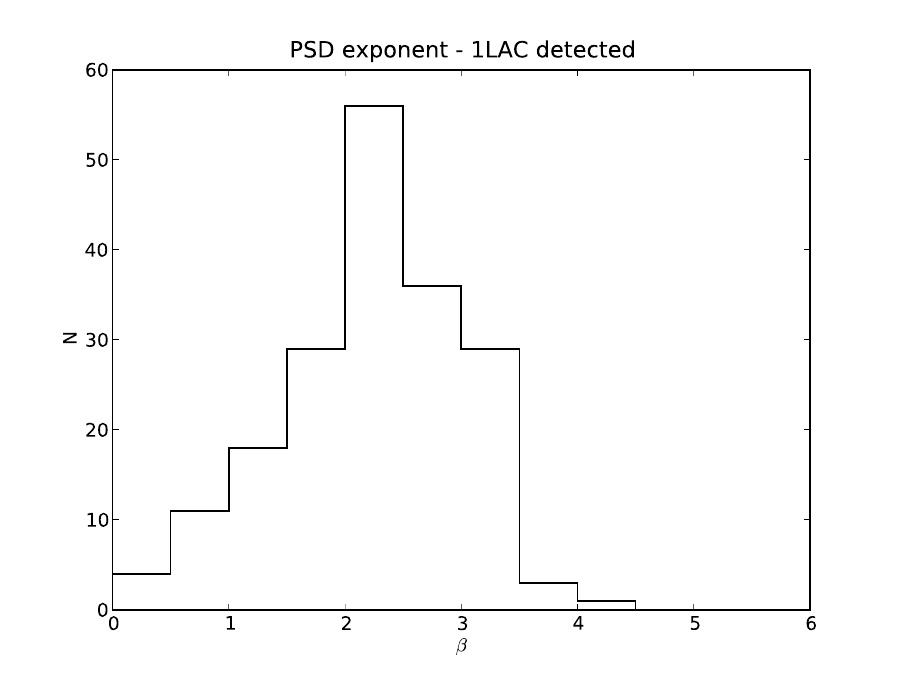}
\includegraphics[width=8cm]{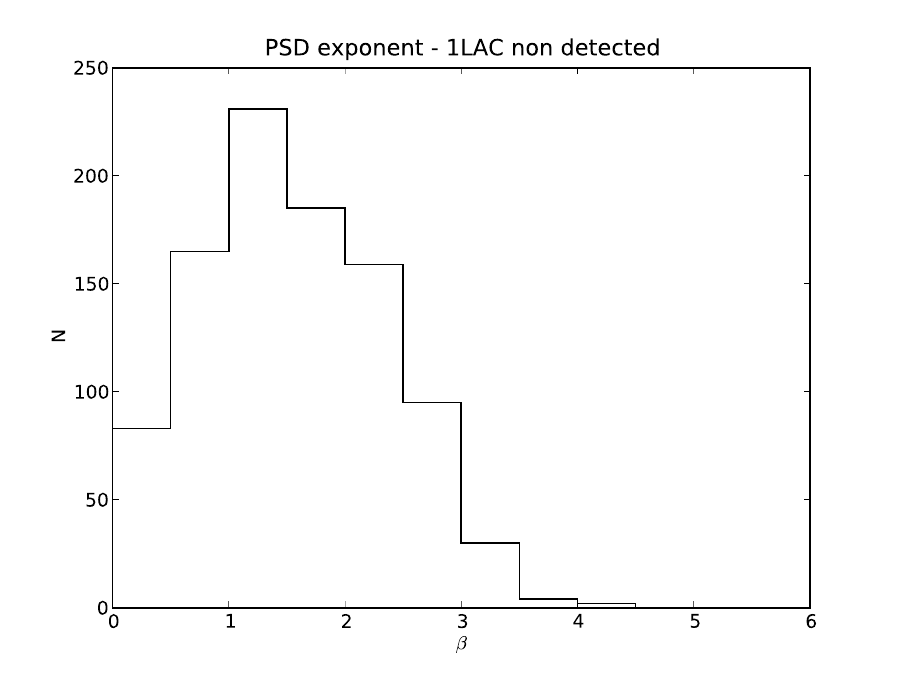}
\caption{Comparison of the power law index of the PSD for gamma-ray detected sources versus the non-detected in our radio sample. Upper panel is the distribution of power law indices for the ones detected in the first \emph{Fermi} AGN catalog. Lower panel is the distribution for the non-detected sources in our sample.} \label{quietvsloudpsd}
\end{figure}

The power spectral density (PSD) is an excellent tool to characterize the variability of light curves using the time domain information. Besides providing another dimension in variability studies, it is key for an accurate estimation of the statistical significance of the cross-correlations using the methods discussed in the previous section. Current gamma-ray light curves only provide enough information to study this on the brightest sources. This has already been done and presented in \citet[][]{abdo_variability_2010} using evenly sampled light curves for a frequency range going from $\sim$0.003 day$^{-1}$ to $\sim$0.1 day$^{-1}$. In the case of radio data we have a larger set of time series, and the main challenge is a robust treatment of unevenly sampled data. Our data set allow us to study the PSD in the frequency range going from $\sim$0.002 day$^{-1}$ to $\sim$0.1 day$^{-1}$. We have adapted a Monte Carlo technique developed for X-ray studies and presented in \citet[][]{uttley_2002}. In their approach the effects of uneven sampling are incorporated by fitting the observed data to simulated light curves with a test power spectral density model. Several models are tested at the same time and a $\chi^2$-like test is used to select the best representation of the data. X-ray light curves have many more data points than a typical radio light curve so the constraints we can obtain are not as strict as what is found in their studies. The situation will improve as we collect longer time series. For the moment we can start exploring population properties by looking at the distributions of power spectral densities for different groups of sources. As an example of what we can learn with these studies we present a comparison of the power law index of the power spectral density in radio for gamma-ray detected sources versus the non-detected (Fig. \ref{quietvsloudpsd}). 
This preliminary analysis indicates that gamma-ray detected sources have radio power spectral densities that are steeper than for the gamma-ray non-detected. The main difference between steep and flat power spectral densities is that steeper ones show more features that look like flares in the time domain.

Further investigation of the power spectral density of radio and gamma-ray light curves can provide more clues about what characterizes gamma-ray loud sources and also enable more accurate estimates for the statistical significance of cross-correlations.

\section{Future Developments}
Our radio monitoring program will continue for at least several more years. With a longer time base available we will be able to improve the estimates of the significance by using light curves with a larger number of episodes of high activity. As \emph{Fermi} detects more sources we can also improve our understanding of population properties as seen from the radio emission. A number of projects aimed at taking advantage of this monitoring program are underway. A brief description of them is given below.

\subsection{Extending Cross-correlation Studies}
A project to combine gamma-ray light curves for the first 3 years of \emph{Fermi} observations is underway. The longer time series and a larger sample of sources will provide more information about the relationship of the radio and gamma-ray emission in blazars and its variations across different source populations. The longer time series will also allow us to better constrain the power law index of the PSDs which will also improve constraints on the significance of the cross-correlation.

\subsection{Radio Polarization Monitoring}
Radio emission in blazars is produced by the synchrotron mechanism and it is polarized. Variations in the polarization can help us understand the evolution of the magnetic fields in the emission regions and its relation to high-energy emission. We are currently only measuring continuum flux but we are building a new receiver which will also measure linear polarization in the Ku-band. This new receiver, which we call KuPol, will cover the band from 12 GHz to 18 GHz with 16 MHz spectral resolution, which can be used for examining spectral indices and removal of radio frequency interference. This will be an important addition to our program and commissioning is planned for the summer of 2012.

\section{SUMMARY}
Our radio monitoring program using the OVRO 40 meter Telescope will soon enter its fifth year. We are currently monitoring more than 1700 sources continuously. Unbiased and continuous monitoring has allowed us to find important differences between the radio characteristics of gamma-ray detected sources versus the non-detected population, such as the larger modulation index for gamma-ray detected sources and the preliminary indication for steeper power spectral densities for this same population. We have introduced the use of robust statistical techniques for the study of variability and its correlation among different wavebands. Using this method with values of the power spectral density power law of -2.0 at 15 GHz and -1.5 for gamma-ray energies, 7 of our 52 objects show cross-correlations at the $3\sigma$ level. We are now studying the physical significance of these correlations by further exploring the range of power law slopes that are consistent with the data. We are currently working on augmenting the sizes of the populations being studied and the time span of the observations. We are also updating the 40 meter telescope receiver to include polarization monitoring which will allow a study of the evolution of magnetic fields in the emission regions.

\bigskip 
\begin{acknowledgments}
The OVRO 40 M program is supported in part by NASA grant NNX11AO43G and NSF grant AST-1109911. VP acknowledges support for this work provided by NASA through Einstein Postdoctoral Fellowship grant number PF8-90060 awarded by the Chandra X-ray Center, which is operated by the Smithsonian Astrophysical Observatory for NASA under contract NAS8-03060.
\end{acknowledgments}

\bigskip 

\end{document}